\documentclass[10pt]{article}
\usepackage{amssymb,pstcol,times,color,graphicx,graphics,pstricks,pst-node,pst-coil,mathrsfs,amsfonts,colortbl,fancyhdr}
\begin{document}
\date{}
\title{Effective action in DSR1 quantum field theory}
\author
{M V Cougo-Pinto$^{a}$\footnote{marcus@if.ufrj.br}, C Farina$^{a}$\footnote{farina@if.ufrj.br}, J F M Mendes$^{b}$\footnote{jayme@ime.eb.br}, F S S Rosa$^{a}$\footnote{siqueira@if.ufrj.br}\\
\\
$^{a}${\it UFRJ, Instituto de F\' \i sica, CP 68528, Rio de Janeiro, RJ,
21.941-972}\\
$^{b}${\it IME, Rio de Janeiro, RJ,
22.290-270}
\\}
\maketitle
\begin{abstract}
We present the one-loop effective action
of a quantum scalar field with DSR1 space-time symmetry as a sum over
field modes. The effective
action has real and imaginary parts and manifest charge conjugation 
asymmetry, which provides an alternative theoretical setting to the study 
of the particle-antiparticle asymmetry in nature. 
\end{abstract}

We report here a recent result concerning the one-loop effective action
of doubly-relativistic theory of type 1 (DSR1) \cite{BrunoAmelino-CameliaKowalski-Glikman01,Amelino-Camelia02Review}. 
To set the stage for this result we start by discussing  
analogous results in usual relativistic theory described by the Poincar\'e
algebra. Then we consider the $\kappa$-deformed Poincar\'e algebra in
standard basis and in bicrossproduct basis  \cite{LukierskiNowickiRueggTolstoy91,LukierskiNowickiRuegg92,
MajidRuegg94,LukierskiRueggZakrzewski95}. The latter is the mathematical
setting of DSR1 theory \cite{Amelino-Camelia02Review}.

It is usually assumed that local space-time symmetries are completely 
described by the Poincar\'e algebra, whose defining commutation 
relations between the basis elements we exhibit here with the purpose 
of comparison with the deformed algebras that we consider in the present work. In usual notation we have
\begin{eqnarray}\label{Poincare}
\begin{array}{c}
[P^{\mu},P^{\nu}] = 0 \; , \\
\vspace{-0.40cm}\\
i[P^{\mu},J^{\rho \sigma}]=g^{\mu\rho} P^{\sigma}-g^{\mu\sigma} P^{\rho}
\; ,\\
\vspace{-0.40cm}\\
{i [ J^{\mu \nu}, J^{\rho \sigma} ]} = {g^{\nu \rho} J^{\mu \sigma} - 
g^{\mu \rho}J^{\nu \sigma} - g^{\nu \sigma} J^{\mu \rho} 
+ g^{\mu \sigma}J^{\nu \rho}} \; ,  \\
\end{array}
\end{eqnarray}
where we identify $P^0$ as the energy, $P^i$ ($i=1,2,3$) as the 
components of momentum three-vector, $J^i=\varepsilon^{ijk}J^{jk}/2$ 
($i=1,2,3$) as the components of angular-momentum three-vector and 
$K^i=J^{i0}$ ($i=1,2,3$) as the components of the boost three-vector.
The first Casimir invariant of this algebra is the relation between
energy and momentum given by
\begin{equation}\label{Casimir}
P_{\mu}P^{\mu}={\bf P}^2-P_0^2=-m^2 \; ,
\end{equation}
where $m^2$ is a positive real scalar which labels the irreducible
representation of the algebra under consideration. For this
case $m$ is the mass of the field excitation, {\it i.e.}, its rest 
frame energy. It leads to the usual relativistic dispersion relations
$p^0=\omega^{(\pm)}({\bf p})$, where
\begin{equation}\label{usualdispersion}
\omega^{(\pm)}({\bf p})=\pm\sqrt{{\bf p}^2+m^2} \; .
\end{equation}
The equality $\omega^{(+)}({\bf p})=-\omega^{(-)}({\bf p})$ describes
charge conjugation symmetry, {\it i.e}, particles and antiparticles
have the same dispersion relation (except for the signs).

From the Casimir invariant (\ref{Casimir}) we also identify the Schwinger 
proper-time Hamiltonian for a relativistic scalar field \cite{Schwinger51},
\begin{equation}\label{H}
H={\bf P}^2-P_0^2+m^2 \; .
\end{equation}
For simplicity we will be considering here only the case of a scalar field,
which we represent by $\phi$.

The proper-time Hamiltonian is all we need in order to calculate
the one-loop effective action given the in Schwinger
representation, which has the form \cite{Schwinger51}
\begin{equation}\label{acaotempoproprio}
\mathcal{W} = -\frac{i}{2} \int_{0}^{\infty} \frac{ds}{s} \;
Tr\; e^{-isH}
\end{equation}
where the real variable $s$ is the so-called proper time, H is
the proper-time Hamiltonian of the theory and $Tr$ represents the total trace on
the exponential operator. The one-loop effective action $\mathcal{W}$
satisfies
\begin{equation}\label{acaoefetivaVEV}
\langle 0\,{t_2} \vert 0\,{t_1}
\rangle  = e^{i \mathcal{W}}
\end{equation}
where $\langle 0\,{t_2} \vert 0\,{t_1}
\rangle$ is the vacuum persistence amplitude
from an instant $t_1$ in the remote past to an
instant $t_2$ in the distant future.
In (\ref{acaoefetivaVEV}) we identify the vacuum energy $E_0$
as given by the real part of the effective action,
\begin{equation}\label{vacuumenergy}
E_0=-\frac{\Re \mathcal{W}}{T} \; ,
\end{equation}
where $T=t_2-t_1$. The shift of this energy caused by external conditions
is the Casimir energy \cite{Casimir48,BordagMohideenMostepanenko01}, 
an important observable in any relativistic
field theory. The properly renormalized imaginary part of the effective 
action gives the probability of vacuum decay,
\begin{equation}\label{moduloVEV}
1 - \vert \langle 0_{t_2} \vert 0_{t_1}
\rangle \vert^{2}  = 1 - e^{- 2 \Im \mathcal{W}} \; .
\end{equation}
In this way, a positive imaginary contribution coming from the
effective action may give rise to the phenomenon of creation of
field excitations. Obviously enough, a negative imaginary part
would lead to the inconsistent result of a negative probability.
It is remarkable that two genuine and important quantum field 
effects, such as the Casimir energy and vacuum decay, can be obtained
from the sole information of what is the Casimir invariant of
the algebra of space-time symmetries of the field. From this 
algebraic invariant we also obtain the expression of the vacuum
energy as the half sum of field frequencies. Indeed, by substituting 
into (\ref{acaotempoproprio}) the proper-time Hamiltonian $H$ given 
in (\ref{H}) we arrive at
\begin{equation}\label{W}
\frac{\mathcal{W}}{T}=
-\sum_{\bf p}\frac{1}{2}|\omega^{(\pm)}({\bf p})|\; ,
\end{equation}
where $\omega^{(\pm)}({\bf p})$ is given by the usual dispersion 
relations (\ref{usualdispersion}). The identification of the half
sum of frequencies in (\ref{W}) with the vacuum energy then follows
from the expression (\ref{vacuumenergy}) of the effective action
in terms of the vacuum energy. 

Results in quantum gravity and string theory, as well as some
experimental astrophysical paradoxes, seem to indicate the existence
of new dispersion relations which differ from the usual relativistic relations (\ref{usualdispersion}) \cite{Amelino-Camelia02Review,Amelino-Camelia02}. 
Those new relations depend on a dimensionful parameter that set the scale
at which the usual relation fails and the new ones should be considered.
In a certain limit of the parameter the new relations reduce to the usual
one, due to the fact that at this limit we are back to the 
regime of validity of Poicar\'e invariance. In this way the new relations 
can be mathematically described as deformations of the usual one and the dimensionful parameter plays the role of a deformation parameter. The proper way of obtaining those deformed dispersion relations is from Casimir invariants
of deformations of Poincar\'e algebras. We are going to deal with
the $\kappa$-deformed Poincar\'e algebras \cite{LukierskiNowickiRueggTolstoy91,LukierskiNowickiRuegg92,
MajidRuegg94,LukierskiRueggZakrzewski95}, as the most promising mathematical
structure to describe possible new space-time symmetries. Those $\kappa$-deformed Poincar\'e algebras are examples of Hopf algebras
also known in the literature as quantum algebras or quantum groups
(cf., e.g, \cite{ChaichianDemichev96}). Since we are not concerned with the co-algebraic part of such quantum groups here we exhibit only its algebraic
parts.

The original $\kappa$-deformed Poincar\'e algebra \cite{LukierskiNowickiRueggTolstoy91,LukierskiNowickiRuegg92}
is defined in the so called standard basis
\cite{LukierskiRueggZakrzewski95} and is given by
\begin{eqnarray}\label{kappaPoincare}
\begin{array}{c}
\left[P^\mu,P^\nu\right]=0 \; ,\\
\vspace{-0.25cm}\\
\left[J^i,J^j\right]=i\,\varepsilon^{ijk}\,J^k\;,
\hspace{20pt}
\left[J^i,P^j\right]=i\,\varepsilon^{ijk}\,P^k\; ,
\hspace{20pt}
\left[J^i,P^0\right]=0\; ,\\
\vspace{-0.25cm}\\
\left[K^i,K^j\right]=
-i\,\varepsilon^{ijk}\,
\left(J^k\,\cosh({P^0}/{\kappa})-({P^k}/{4\kappa^2})\,{\bf P}\cdot{\bf J}\right) \; , \\
\vspace{-0.25cm}\\
\left[K^i,P^j\right]=i\,\delta^{ij}\,\kappa\, \sinh({P^0}/{\kappa})\; ,
\\
\vspace{-0.25cm}\\
\left[K^i,P^0\right]=i\,P^i\; ,
\hspace{20pt}
\left[J^i,K^j\right]=-i\,\varepsilon^{ijk}\,K^k\; ,
\end{array}
\end{eqnarray}
where $\kappa$ is a real positive deformation parameter with
dimension of mass. This parameter sets for energy, length and time the
respective scales $E_{\kappa}=\kappa c^2$, $\ell_{\kappa}=\hbar/(\kappa c)$ and
$t_{\kappa}=\hbar/(\kappa c^2)$. The Casimir invariant relating energy and 
momentum in this algebra is given by 
\begin{equation}\label{kappaCasimir}
{\bf P}^2-\left(2\kappa\,\hbox{sinh}\frac{P^0}{2\kappa}\right)^2=-m^2 \; ,
\end{equation}
which leads immediately to the following dispersion relations
\begin{equation}\label{kappadispersion}
p^0=\omega^{(\pm)}({\bf p})=\pm
2\kappa\,\hbox{sinh}^{-1}\left(\frac{1}{2\kappa}\sqrt{{\bf
p}^2+m^2}\right)\, .
\end{equation}
In this theory, that we call for short $\kappa$-deformed theory, the
proper-time Hamiltonian is
\begin{equation}\label{Hkappa}
H_{\kappa}={\bf P}^2 -(2\kappa)^2{\rm
sinh}^2\left(\frac{P_o}{2\kappa}\right)+m^2 \; .
\end{equation}
In the limit in which the deformation parameter $\kappa$ goes to infinity
the $\kappa$-deformed Poincar\'e algebra (\ref{kappaPoincare}) reduces
to the usual Poincar\'e algebra (\ref{Poincare}). In this limit the $\kappa$-deformed Casimir invariant (\ref{kappaCasimir}), dispersion 
relations (\ref{kappadispersion}) and proper-time Hamiltonian (\ref{Hkappa})
also reduce to the corresponding non-deformed quantities given in 
(\ref{Casimir}), (\ref{usualdispersion}) and (\ref{H}), respectively.

By inserting this $\kappa$-deformed proper time Hamiltonian (\ref{Hkappa})
into (\ref{acaotempoproprio}), we obtain the following expression 
for the $\kappa$-deformed effective action
\begin{equation}\label{kappaW}
\frac{\mathcal{W}_{\kappa}}{T}=
-\sum_{\bf p}\frac{1}{2}|\omega^{(\pm)}({\bf p})|
+i\frac{1}{2\pi\kappa}\sum_{\bf p}\frac{1}{2}|\omega^{(\pm)}({\bf p})|^2 \; ,
\end{equation}
where now $\omega^{(\pm)}({\bf p})$ is obtained from the $\kappa$-deformed
dispersion relation (\ref{kappadispersion}). The real part has the
usual form of half sum of field frequencies as in the non-deformed case, although the frequencies which are summed here are the $\kappa$-deformed     (\ref{kappadispersion}). The imaginary part has the interesting property 
of being proportional to the sum of the squares of the $\kappa$-deformed 
field frequencies. If the scalar field is submitted to 
Dirichlet boundary conditions on parallel plates
of side $\ell$ and separation $a$, we obtain from the real part of the $\kappa$-deformed effective action the Casimir energy \cite{C-PFarina97}
\begin{equation}\label{kappaCasimirenergy}
{\cal E}(a)=-\frac{\ell^2}{16\pi^2a^3}
\sum_{n=1}\int_{0}^{\infty} d\sigma \sigma
e^{-n^2\sigma-(2\kappa^2+m^2)a^2/\sigma}
\sqrt{\frac{4a^2\kappa^2}{\pi\sigma}}
\pi{\rm I}_0\left(\frac{2a^2\kappa^2}{\sigma}\right)
\end{equation}
and from the imaginary part the following creation rate of field excitations \cite{C-PFarina97}
\begin{equation}\label{kappaCreation}
\frac{\Im \mathcal{W}_{\kappa}}{T a\ell^2}=\frac{1}{16\pi^2a^4}
\sum_{n=1}\int_{0}^{\infty} d\sigma \sigma
e^{-n^2\sigma-(2\kappa^2+m^2)a^2/\sigma}
\sqrt{\frac{4a^2\kappa^2}{\pi\sigma}}
{\rm K}_0\left(\frac{2a^2\kappa^2}{\sigma}\right) \; ,
\end{equation}
where ${\rm I}_0$ and ${\rm K}_0$ are the modified Bessel 
functions. It is important to note the manifest positiveness of the 
renormalized imaginary part (\ref{kappaCreation}) of the effective action. 
The $\kappa$-deformed Casimir energy (\ref{kappaCasimir})
reduces to the Casimir energy of the usual scalar field
in the limit $\kappa\rightarrow\infty$ in which the deformation
disappears. The creation rate goes to zero in this limit or
in the limit $a\rightarrow\infty$ of infinite separation of 
the plates. In this way the creation rate is a consequence of
both deformation and boundary condition; if one of those factors
is not present the creation rate vanishes. 

Let us now turn to the $\kappa$-deformation of the Poincar\'e algebra 
in the so called bicrossproduct basis, which is given by 
\cite{MajidRuegg94,LukierskiRueggZakrzewski95}
\begin{eqnarray}\label{algebraDSR}
\begin{array}{c}
[P_{\mu},P_{\nu}] = 0  \\
\vspace{-0.05in} \\
{ [J_i, P_j]} = i \epsilon_{ijk} P_k   \;\;\;\;\;\;\;\;\;\;\;  [J_i, P_0]
= 0 \\
\vspace{-0.05in} \\
{[K_i, P_j]} = i \delta_{ij} 
\left((1/2\lambda)(1 - e^{2 P_0\lambda})
+ (\lambda/2) {\bf P}^2 \right) - i \lambda P_iP_j \\
\vspace{-0.05in} \\
{ [K_i, P_0]} = i P_i \\
\vspace{-0.07in} \\
{i [ J_{\mu \nu}, J_{\rho \sigma} ]} = {g_{\nu \rho} J_{\mu \sigma} - g_{\mu \rho}
J_{\nu \sigma} - g_{\nu \sigma} J_{\mu \rho} + g_{\mu \sigma}J_{\nu \rho}}  \\
\end{array}
\end{eqnarray}
where $\lambda=1/\kappa$ is an alternative parameter with the dimension of length. The doubly special relativity of type 1 is a theory whose 
space-time symmetry is described by this algebra \cite{BrunoAmelino-CameliaKowalski-Glikman01,Amelino-Camelia02Review}.
This parameter provides the second invariant, besides the speed of light,
which characterizes the theory as doubly special relativity. In DSR1 there is
a natural scale for length $\ell_\lambda=\lambda$ and also scales for energy and time, namely $E_\lambda=\hbar c/\lambda$ and $t_\lambda=\lambda/c$. In most applications $\ell_\lambda$ is taken as the Planck length. For brevity we
refer to (\ref{algebraDSR}) as the $\lambda$-deformed Poincar\'e algebra.
In the limit $\lambda\rightarrow 0$ this deformed algebra reduces to the Poincar\'e algebra.

From the $\lambda$-deformed Poincar\'e algebra (\ref{algebraDSR}) we have
the following Casimir invariant involving energy and momentum
\begin{equation}\label{DSR1Casimir}
{\bf P}^2 e^{\lambda P_0}-
\frac{e^{\lambda P_0}+ e^{-\lambda P_0} - 2}{\lambda^2}=-m^2 \; ,
\end{equation}
Let us take from this Casimir invariant the following proper-time Hamiltonian for a scalar field
\begin{equation}\label{Hlambdastar}
H_{\lambda*} = {\bf P}^2 e^{ \lambda P^0} + m^2 -
\frac{1}{\lambda^2} \left( e^{ \lambda P^0} - 2  + e^{- \lambda P^0}
\right).
\end{equation}
From the Casimir invariant (\ref{DSR1Casimir}) we also identify the 
$\lambda$-deformed dispersion relations $p^0=\omega^{(\pm)}({\bf p})$, 
where
\begin{eqnarray}\label{positivefrequency}
\omega^{(+)}({\bf p})=-\frac{1}{\lambda}
\log\left[1+\frac{\lambda^2 m^2}{2}
-\sqrt{\left(\frac{\lambda^2 m^2}{2}\right)^2
+\lambda^2\left({\bf p}^2+m^2\right)}\right]
\end{eqnarray}
are positive frequencies defined for $|{\bf p}|<1/\lambda$ and
\begin{eqnarray}\label{negativefrequency}
\omega^{(-)}({\bf p})=-\frac{1}{\lambda}
\log\left[1+\frac{\lambda^2 m^2}{2}
+\sqrt{\left(\frac{\lambda^2 m^2}{2}\right)^2
+\lambda^2\left({\bf p}^2+m^2\right)}\right]
\end{eqnarray}
are negative frequencies defined for any $|{\bf p}|$. Notice 
that for $|{\bf p}|>1/\lambda$ expression (\ref{positivefrequency}) 
for the positive frequency is not a real number and for this reason
it should be discarded. However, we will see that its real part 
\begin{eqnarray}\label{transpositive}
\omega^{(>)}({\bf p})=-\frac{1}{\lambda}
\log\left[-1-\frac{\lambda^2 m^2}{2}
+\sqrt{\left(\frac{\lambda^2 m^2}{2}\right)^2
+\lambda^2\left({\bf p}^2+m^2\right)}\right] \; ,
\end{eqnarray}
for $|{\bf p}|>1/\lambda$, appears naturally in the development 
of the formalism due to the fact that the 
Casimir invariant (\ref{DSR1Casimir}) was taken unrestrictedly to define 
the proper-time Hamiltonian (\ref{Hlambdastar}) of the theory. 
The theory arising from those dispersion relations has no invariance under
charge conjugation. In particular, particles have the saturation
momentum $p_c = 1/\lambda$ while antiparticles have no constraint on
its momentum values. Let us call the momentum or any quantity 
thereof depending unsaturated or transaturated whenever the magnitude
of the momentum is lower or greater than the saturation momentum $1/\lambda$,
respectively.  

As an alternative way of defining a theory from the Casimir invariant (\ref{DSR1Casimir}) it is possible to use only the dispersion
relation $p^0=\pm\omega^{(+)}({\bf p})$ with charge conjugation symmetry 
imposed on the theory. In this case both particles and antiparticles
have saturation momentum $1/\lambda$. We will consider the theory 
with charge conjugation asymmetry derived from the proper-time Hamiltonian
(\ref{Hlambdastar}). To distinguish it from other possibilities we refer to 
it as DSR1$^*$ theory. The main result that we present for the DSR1$^*$ theory
is a calculation of its one-loop effective action leading to an expression
in terms of sum of field modes. 

The calculation of the effective action for DSR1$^*$ theory is a
rather involved problem, due to its most essential features, namely:
the lack of charge conjugation symmetry in the theory and the
existence of a saturation momentum $1/\lambda$, which leads to a
rather peculiar distribution of poles in the plane of complex
frequencies. For simplicity we consider a two-dimensional
space-time. Although not necessary for our purposes
here, it is convenient for future applications to
impose periodic boundary conditions on the scalar field $\phi$,
$\phi(x + a,t) = \phi(x,t)$, where $a$ is the periodicity length.
In this case momentum is discretized as $p^1={2 \pi n}/{a}$
($n \in \mathbb{Z}$) and the saturation momentum becomes
$[a/(2\pi\lambda)]$, the greatest integer which is smaller than
$a/(2\pi\lambda)$. After a long calculation which starts by
substituting the DSR1$^*$ proper-time Hamiltonian (\ref{Hlambdastar})
in the Schwinger's formula (\ref{acaotempoproprio}) we obtain the
following expression for the one-loop DSR1$^*$ effective action
\begin{eqnarray}\label{acao efetiva final}
\frac{\mathcal{W}_{\lambda}}{T}\hspace{-17pt}&&=
-\frac{1}{2}\sum_{\vert n\vert\leq\left[\frac{a}{2\pi\lambda}\right]}
\vert \omega_{n}^{(-)} \vert-
\frac{1}{4}\sum_{\vert n\vert>\left[\frac{a}{2\pi\lambda}\right]}
\vert \omega_{n}^{(-)} \vert
\nonumber\\
&&+i\; \frac{\lambda}{2 \pi}
\!\left[\frac{1}{2}
\sum_{n \in \mathbb{Z}} \!\!\left( \omega_{n}^{(-)} \right)^2 +
\frac{1}{2}
\!\!\sum_{\vert n \vert \leq \left[ \frac{a}{2 \pi\lambda} \right]}
\left( \omega_{n}^{(+)} \right)^2
+ \frac{1}{2}
\!\!\sum_{\vert n \vert > \left[ \frac{a}{2 \pi\lambda} \right]}
\!\!\left( \omega_{n}^{(>)}
\right)^2 \right]\! ,
\end{eqnarray}
where $\omega_{n}^{(+)}$, $\omega_{n}^{(-)}$ and $\omega_{n}^{(>)}$
are the discretized versions of the frequencies defined in (\ref{positivefrequency}), (\ref{negativefrequency}) and
(\ref{transpositive}), respectively. By comparing this expression
for the DSR1$^*$ action with the corresponding expression
(\ref{kappaW}) for the $\kappa$-deformed action it is evident 
that the former has a more complicated structure that seems to be a 
result of the asymmetry between the particle and antiparticle 
sectors and of the existence of a saturation momentum. 
The expression (\ref{acao efetiva final}) for the effective action 
as a sum of modes is well suited for physical interpretation, to which we proceed now.

First of all, the DSR1$^*$ effective action has real and imaginary parts, 
as in the $\kappa$-deformed case, and the whole expression has the correct 
limit when the deformation disappears, {\it i.e.}, when $\lambda\rightarrow 0$.
In this limit the imaginary part of the DSR1$^*$ goes to zero, because of 
its prefactor $\lambda$, while the real part reduces to the effective 
action (\ref{W}) of the non-deformed scalar field. Also as
in the $\kappa$-deformed case the real part is given by a sum of 
field frequencies and the imaginary part by a sum of squares of 
field frequencies. We are led to raise the question whether any
deformation of the Poincar\'e algebra will have both real and imaginary
parts and, if this is the case, it will follow the pattern
of sum of frequencies for the real part and sum of squares of frequencies
for the imaginary part. 

Due to lack of charge conjugation symmetry in DSR1$^*$ theory its effective 
action reveals that stationary vacuum fluctuations obey the dispersion
relations of the negative frequencies while the possible vacuum decay 
described by the imaginary part of the action occurs according to 
the dispersion relations of both negative and positive frequencies.
Clearly, those features cannot be seen nor in the non-deformed theory 
neither in the $\kappa$-deformed theory, because the negative 
and positive frequencies in their effective actions have the same modulus, 
due to charge conjugation symmetry. Besides the sole presence of negative frequencies in the real part, they appear in a rather peculiar way, namely: the unsaturated frequencies with the usual weight $1/2$ in its sum and the transaturated frequencies with just half of this weight, $1/4$.

We should also notice that in the imaginary part of the
effective action there are contributions not only from the positive and
negative frequencies, but also from the frequencies $\omega_{n}^{(>)}$
that we defined in (\ref{transpositive}). It is remarkable that the real 
quantity $\omega_{n}^{(>)}$ is not included into the theory by hand, but appears naturally in the course of calculation. With the presence of those transaturated positive frequencies in the imaginary part we may say that the whole range of 
momenta contributes to the imaginary part, with the sole exception of
the saturation momentum $1/\lambda$. 

Due to the presence of both real and imaginary parts in DSR1$^*$ theory 
it is open the possibility that the real part gives rise to
a Casimir energy when boundary conditions are present and the imaginary
part to a creation rate of field excitations even with static boundary conditions. However, it is not a straightforward task the calculation 
of those quantities, due to the abrupt cut in the summation of frequencies
when saturation moment is reached, as it happens in the real and imaginary
parts of DSR1$^*$ effective action (\ref{acao efetiva final}). This is
a problem that deserves careful analysis, since Casimir energy and
creation rates associated to boundary conditions are useful
theoretical probes into any theory. For example, a negative imaginary
part in the renormalized effective action reveals a serious 
inconsistency in the theory, as remarked above. 

Finally, we should make a remark about the lack of charge
conjugation symmetry in DSR1$^*$. Indeed, the overwhelming asymmetry
between the quantities of matter and antimatter in the universe
requires an asymmetry somewhere in the theory of its evolution.
It has been argued \cite{C-PFarina97} that the creation rate of field excitations (\ref{kappaCreation}) obtained from the imaginary part of the
$\kappa$-deformed effective action provides a mechanism for the
creation of matter and radiation in the early universe. Now, if
a theory has not only an imaginary part in the effective action,
but also a manifest charge conjugation asymmetry, it is open the 
possibility that the above mentioned mechanism of creation of matter 
and radiation occurs in a non-symmetrical way. This may provide an
alternative theoretical setting to investigate the particle-antiparticle
asymmetry in nature. As we have seen, DSR1$^*$ is an example of a 
theory with those features, although we cannot yet decide for sure
what would be the correct theory describing such asymmetry in nature.
Much more work is necessary to reach any conclusion on such matters,
but it is worthwhile to consider the above remarks, due to the importance 
of the subject.

There are some natural continuations to this work. The obvious one
is to calculate the Casimir energy and possible creation rate of field
excitations from the DSR1$^*$ effective action. The result of this
calculation may be very important to determine the consistency of the
theory and its general features as, for example, if the presence of
transaturated quantities in the effective action is spurious or not.
If they are spurious we will have to face the question of how to 
eliminate them.

\noindent
{\bf Acknowledgments}\\
C.F. and F. S. S. R. acknowledge CNPq for partial financial support.
\end{document}